\documentclass[twocolumn,showpacs,preprintnumbers,amsmath,amssymb]{revtex4}

\usepackage{epsfig}
\usepackage{graphics}
\usepackage{graphicx}
\usepackage{dcolumn}
\usepackage{bm}

\begin{document}


\title{Transition to complete synchronization and global intermittent synchronization in an array of time-delay systems}

\author{R.~Suresh$^1$}
\author{D.~V.~Senthilkumar$^{2}$}
\author{M.~Lakshmanan$^1$}
\author{J.~Kurths$^{2,3,4}$}

\affiliation{$^1$Centre for Nonlinear Dynamics, School of Physics, Bharathidasan University, Tiruchirapalli 620 024, India\\
$^2$Potsdam Institute for Climate Impact Research, 14473 Potsdam, Germany\\
$^3$Institute of Physics, Humboldt University, 12489 Berlin, Germany\\
$^4$Institute for Complex Systems and Mathematical Biology, University of Aberdeen, Aberdeen AB24 3UE, United Kingdom}
\date{\today}

\begin{abstract}
We report the nature of transitions from nonsynchronous to complete synchronization (CS) state in arrays of time-delay
systems, where the systems are coupled with instantaneous diffusive coupling. 
We demonstrate that the transition to CS occurs distinctly for different coupling
configurations. In particular, for unidirectional coupling, locally (microscopically) synchronization transition occurs 
in a very narrow range of coupling strength but for a global one (macroscopically) it occurs 
sequentially in a broad range of coupling strength preceded by an intermittent synchronization. 
On the other hand, in the case of mutual coupling a very large value of coupling strength is required for local synchronization 
and, consequently, all the local subsystems synchronize 
immediately for the same value of the coupling strength and hence globally synchronization also occurs
in a narrow range of the coupling strength. In the transition regime, we observe a new type of 
synchronization transition where long intervals of high quality synchronization 
which are interrupted at irregular times by intermittent chaotic bursts simultaneously  
in all the systems, which we designate as {\it global intermittent synchronization (GIS)}.
We also relate our synchronization transition results to the above specific types using unstable periodic orbit theory. 
The above studies are carried out in a well known piecewise linear time-delay system.
\end{abstract}

\maketitle

\section{\label{sec:level1}Introduction}
Numerical and experimental investigations of chaotic synchronization in coupled nonlinear
systems have been receiving much attention in recent years. This phenomena is omnipresent
and plays an important role in diverse areas of science and technology \cite{aspmgr2001, sbjk2002}. 
In the synchronization process, two identical chaotic systems do not always necessarily synchronize perfectly. 
Rather, long intervals of high-quality synchronization are interrupted at irregular times by intermittent 
chaotic bursts and such chaotic bursts along with the synchronization are called
on-off intermittency \cite{jfh1994}. It has been shown that on-off
intermittency is a frequently occurring instability
preceding typical synchronization transitions in diverse dynamical systems,
mediated by unstable periodic orbits (UPOs)~\cite {pa1994}. Further, as the coupling parameter is
increased, a periodic orbit embedded in the attractor in the invariant synchronization 
manifold can become unstable for perturbations (such as noise and/or
parameter mismatches) transverse to the manifold. This is called a bubbling
bifurcation, which leads to the formation of riddled basins of attraction 
in the invariant manifold inducing intermittent bursting (see Ref. \cite{sv1996} for more details).
There exists another type of bifurcation, called blowout bifurcation, induced by 
changes in the transverse stability of an infinite number of UPOs. Among these UPOs, 
some are transversely stable and others are transversely unstable near the bifurcation.

It is a well accepted fact that on-off intermittency is a common phenomenon which occurs 
in a wide variety of natural systems, including neural networks \cite{tk2006, jlp1999}, biological systems \cite{ah2006},
laser systems \cite{ms1998, vf2009}, electronic circuits \cite{dlg1996, yhy1995}, 
complex networks \cite{rlv2005}, coupled chaotic systems \cite{hly1996}, 
earthquake occurrence \cite{mb2007} and other physical systems such as Hamiltonian systems 
and self-driven particle systems \cite{jhec2003, ch2004}. Specifically, it has been reported
that the dynamics of clusters in a network can exhibit an extreme form of intermittency \cite{xw2009}:
A substantial percentage of synchronized nodes forms a giant cluster most of the time,
while many small clusters can also occur at other times. Thus the cluster sizes can vary in a highly
intermittent fashion as a function of time.
Recently, it has been shown \cite{lz2005}
that the transition to intermittent chaotic synchronization (in the case of complete synchronization (CS))
for phase-coherent attractors (R$\ddot{o}$ssler attractors) occurs immediately as soon as the 
coupling parameter is increased from zero and for non-phase-coherent attractors (Lorenz attractors)
the transition occurs slowly in the sense that it occurs only when the coupling is sufficiently
strong known as delayed transition.

It has been already shown that the transition from nonsynchronization to any type of 
synchronization is preceded by intermittent synchronization in coupled chaotic systems.
For example, intermittent lag synchronization (ILS) \cite{sb2000}, intermittent phase synchronization (IPS)
\cite{ap1997, kjl1998} and intermittent generalized synchronization (IGS) \cite{aeh2005} are some of the synchronization
transitions characterized by the intermittent behavior as a function of a coupling parameter. 
Recently, IGS has been numerically observed in unidirectionally coupled time-delay systems 
\cite{dvsml2007}. It has been found that the onset of generalized synchronization is
preceded by on-off intermittency and the transition 
behavior is different for different coupling schemes. In particular, the intermittent 
transition occurs in a broad range of coupling strength for error feedback coupling configuration 
and in a narrow range of coupling strength for direct feedback coupling configuration beyond 
certain threshold values of the coupling strength.

The transition between various types of synchronization
and their mechanism are not yet well
understood especially in time-delay systems. Further, the dynamics of a large ensemble of coupled time-delay
systems such as regular and complex networks are not yet well studied and only a very few studies are 
available in the literature \cite{dvsml2011,rsdvsmljk2010}.
The study of synchronization in ensembles of time-delay systems has been receiving central importance
recently in view of the infinite dimensional nature and feasibility of experimental realization of
time-delay systems.
Particularly, considerable attention is being paid to time-delay systems with instantaneous coupling due to their extensive applications in
different fields such as signal and image processing, pattern recognition, chaotic neural
networks, secure communication and cryptography \cite{xw2010, ly2000, gdv1998, aa2005, sb2008, jl2008}.
In particular, In Refs. \cite{gdv1998, aa2005}, it has been demonstrated that in chaotic communication experiments, 
the time-delayed optical fibre ring laser system is capable of transmitting the 
encoded signals with the speed of 1Gb/sec data range over a long distance fibre-optic channel ($\approx 120Km$).

Motivated by the above, we will investigate here synchronization transitions in
an array of coupled time-delay systems with different (instantaneous) coupling configurations.
Particularly, in this paper, we demonstrate that the transition to CS occurs 
distinctly for different coupling configurations in a regular array of coupled time-delay systems. 
In an unidirectional array 
the transition from nonsynchronization to CS occurs locally (microscopically) 
in a narrow range of coupling strength and globally (macroscopically) 
the systems synchronize one by one with 
the drive system as a function of the coupling 
strength, which is known as sequential synchronization. 
But in a mutually coupled array, every individual system synchronizes immediately in a narrow range
(after a large threshold value) of 
the coupling strength and so globally the synchronization transition is immediate as a function of the coupling
strength in contrast to sequential synchronization. It is also to be noted that in the transition
regime we observe a new type of synchronization behaviour called {\it global intermittent synchronization (GIS)}
where long intervals of high quality synchronization are interrupted 
by large desynchronized chaotic bursts simultaneously in all 
the systems in the array.
\begin{figure}
\centering
\includegraphics[width=0.8\columnwidth]{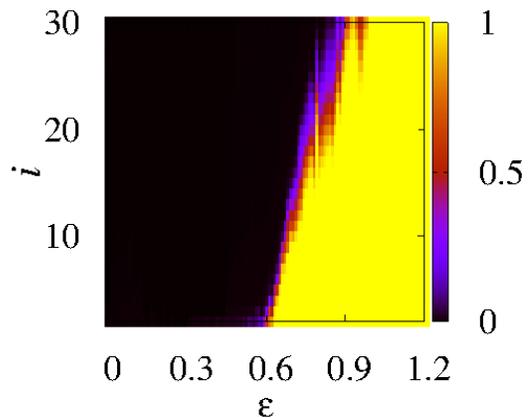}
\caption{\label{fig2} (Color online) The probability of synchronization $\Phi_{i}(\varepsilon)$ as a function 
of the coupling strength $\varepsilon$. The system index $i$ illustrates
the occurrence of sequential synchronization transition to CS in unidirectionally 
coupled piecewise linear time-delay systems (Eq.~(\ref{eqn1})). The black color indicates the
absence of synchronization ($\Phi_{i}(\varepsilon)$=0.0), whereas the yellow color (light gray)
represents the occurrence of CS ($\Phi_{i}(\varepsilon)$=1.0).}
\end{figure}

To understand the two distinct transition scenarios, we focus on the theory of unstable 
periodic orbits, which are the basic building blocks of chaotic and hyperchaotic attractors.
The sequential and immediate synchronization transitions to CS are characterized by calculating the 
probability of synchronization and the average probability of synchronization 
as a function of the coupling strength. The existence of
intermittent synchronization is corroborated by using a spatiotemporal difference and a 
power law behavior of the laminar phase distributions.

The remaining paper is organized as follows: In Sec. \ref{sec:level2}, we will
explain the occurrence of sequential synchronization preceded by intermittent synchronization 
in an array of unidirectionally coupled piecewise 
linear time-delay systems and in Sec. \ref{sec:level4}, we consider a mutual coupling 
configuration and explain the occurrence of instantaneous synchronization transition in 
the array. Further we demonstrate the existence of GIS and provide a possible mechanism for the occurrence of this 
new synchronization transition to CS in the array. Finally, 
we discuss our results and conclusion in Sec.~\ref{sec:level8}.
\section{\label{sec:level2}Synchronization in a piecewise linear time-delay systems: Linear array with unidirectional coupling}
We consider the following unidirectionally coupled time-delay systems of the form
\begin{subequations}
\begin{eqnarray}
\dot{x}_1 &=& -\alpha x_{1}(t)+\beta f(x_{1}(t-\tau)),\\
\dot{x}_i &=& -\alpha x_{i}(t)+\beta f(x_{i}(t-\tau))+\varepsilon[x_{i-1}(t)-x_{i}(t)],
\end{eqnarray}
\label{eqn1}
\end{subequations}
where $i = 2, 3, \cdots, N$. We choose an open end boundary condition. 
$\alpha$, $\beta$ are system parameters, $\tau$ is the time-delay
and $\varepsilon$ is the strength of the coupling between the systems. The nonlinear function
$f(x)$ is chosen to be a piecewise linear function with a threshold nonlinearity, which has 
been studied recently \cite{dvsksmljk2010},
\begin{equation}
f(x) = AF^{*}-Bx.
\label{eqn2}
\end{equation}
Here
\begin{eqnarray}
F^{*}=
\left\{
\begin{array}{cc}
-x^{*},&  x < -x^{*}  \\
            x,&  -x^{*} \leq x \leq x^{*} \\
            x^{*},&  x > x^{*}. \\ 
         \end{array} \right.
\label{eqn3}
\end{eqnarray}

The system parameters for the piecewise linear system (\ref{eqn1})-(\ref{eqn3}) are 
fixed as follows: $\alpha = 1.0$, $\beta=1.2$, $\tau=6.0$, $A=5.2$, $B=3.5$ and
$x^{*}$ is the threshold value fixed at $x^{*}=0.7$. Note that for this set of parameter values a single
uncoupled system exhibits a hyperchaotic attractor with three positive
Lyapunov exponents (LEs) (see Ref. \cite{ks2010}).

\begin{figure}
\centering
\includegraphics[width=0.9\columnwidth]{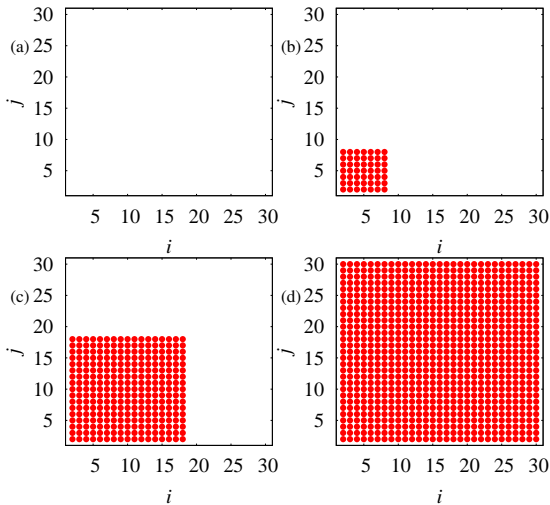}
\caption{\label{fig3e} (Color online) Snap shots of node vs node plots
indicating sequential synchronization in unidirectionally coupled piecewise linear systems 
for different values of coupling strength.
(a) $\varepsilon=0.4$, (b) $\varepsilon=0.7$, (c) $\varepsilon=0.87$ and (d) $\varepsilon=1.1$.}
\end{figure}
\begin{figure}
\centering
\includegraphics[width=0.9\columnwidth]{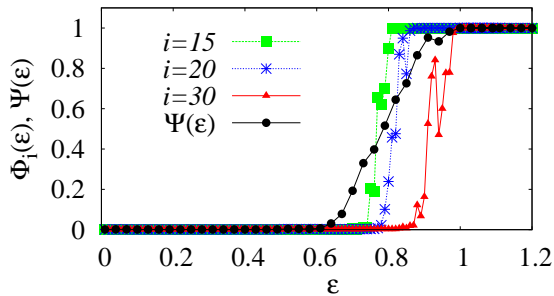}
\caption{\label{fig3} (Color online) The probability of synchronization $\Phi_{i}(\varepsilon)$ of  
selected systems ($i=15,20,30$) and the average probability 
of synchronization ($\Psi(\varepsilon)$) in an unidirectionally coupled array (Eq.~(\ref{eqn1})) 
as a function of the coupling strength $\varepsilon$.}
\end{figure}
\begin{figure*}
\centering
\includegraphics[width=2.1\columnwidth]{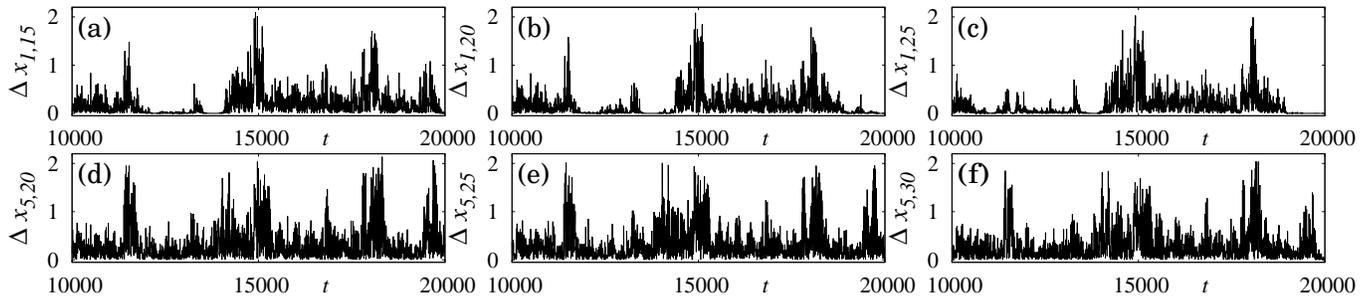}
\caption{\label{fig3a} The difference between some selected 
piecewise linear time-delay systems (Eq.~(\ref{eqn1})) shows intermittent
synchronization. (a) $\Delta x_{1,15}$ for $\varepsilon=0.76$, 
(b) $\Delta x_{1,20}$ for $\varepsilon=0.81$, (c) $\Delta x_{1,25}$ for $\varepsilon=0.86$, 
(d) $\Delta x_{5,20}$ for $\varepsilon=0.76$, (e) $\Delta x_{5,25}$ for $\varepsilon=0.81$ and (f) $\Delta x_{5,30}$ for $\varepsilon=0.86$.}
\end{figure*}
\begin{figure}
\centering
\includegraphics[width=0.8\columnwidth]{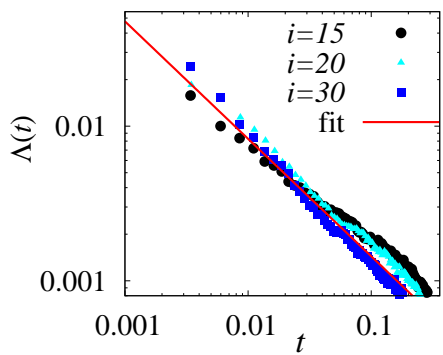}
\caption{\label{fig3b} (Color online) The statistical distribution of the laminar phase for
the systems $i = 15$ for $\varepsilon=0.76$, $i = 20$ for $\varepsilon=0.81$ 
and the system $i=30$ for $\varepsilon=0.93$, all satisfying $-\frac{3}{2}$ power law scaling.}
\end{figure}

To demonstrate the nature of the dynamical transition to a complete synchronization regime, we consider an array of 
$N=30$ unidirectionally coupled identical piecewise linear time-delay systems (\ref{eqn1})-(\ref{eqn3})
(each system having different initial conditions).
Here, $x_{1}(t)$ acts as the drive and the remaining systems ($x_{i}(t), i = 2, 3, \cdots, 30)$ as 
the response systems. In the absence of coupling [$\varepsilon = 0.0$ in Eq.~(\ref{eqn1})] all the 
systems evolve independently according to their own dynamics. On increasing the 
coupling strength, the system $x_{1}(t)$ starts to drive the 
system $x_{2}(t)$. Consequentially, the system $x_{3}(t)$ starts 
to follow the drive system $x_{1}(t)$ for larger values of $\varepsilon$ and this is continued upto the $N^{th}$ system.
Hence, global synchronization is achieved via sequential synchronization
of the systems in the array as a function of coupling strength. To be more specific, 
upon increasing the coupling strength, $\varepsilon$, from zero,
nearby systems to the drive in the array synchronize sequentially with it, while the faraway systems are 
still in their transition state. The other desynchronized systems will synchronize sequentially 
for further larger values of $\varepsilon$.
The occurrence of sequential phase synchronization in an array of unidirectionally
coupled time-delay systems has been shown in Ref.~\cite{rsdvsmljk2010} and
sequential desynchronization in a network of spiking neurons is reported in Ref.~\cite{ck2009} 
as a function of coupling strength, $\varepsilon$. We may also note here a somewhat analogous
situation occurs but now as a function of time for a fixed coupling strength in an array of unidirectionally
coupled chaotically evolving systems in Refs.~\cite{mnl1996, mam1998}.


Here, we find that locally the synchronization in the array occurs immediately
in a very narrow range of coupling strength; globally, it occurs in a broader range of 
$\varepsilon$ due to sequential synchronization. 
To characterize these local and global synchronization transitions, we have calculated the probability
of synchronization $\Phi_{i}(\varepsilon)$ (which is defined as the fraction of time 
during which $\bigl|x_{1}(t)-x_{i}(t)\bigr| < \delta$ occurs, where 
$\delta$ is a small but an arbitrary threshold) and the average probability of synchronization 
[$\Psi(\varepsilon)=\frac{1}{N-1}\sum_{i=2}^{N}\Phi_{i}(\varepsilon)$]. 
Here the asynchronized state is characterized by $\Phi_{i}(\varepsilon) = 0$, CS 
by $\Phi_{i}(\varepsilon) = 1$ and the transition region by intermediate values less than 
unity. 

To understand the dynamical organization of sequential synchronization in the array (Eq.~(\ref{eqn1})),
we have calculated the probability of synchronization as a function of $\varepsilon$
and the system index $i$, which is depicted in Fig.~\ref{fig2}. In this figure, the black 
color indicates the asynchronized state ($\Phi_{i}(\varepsilon)=0.0$) and the yellow color (light gray)
corresponds to the complete synchronization state ($\Phi_{i}(\varepsilon)=1.0$), while intermediate colors
represent the transition region. From this figure one can clearly see the occurrence of 
sequential synchronization as a function of $\varepsilon$ where the nearby
systems to the drive get synchronized first for lower values of $\varepsilon$, whereas
the far away systems are synchronized at larger $\varepsilon$.

Sequential synchronization can also be visualized using snap shots
of the oscillators in the node vs node plots. We regard the oscillators in the array
as synchronized when the probability of synchronization $\Phi_{i}(\varepsilon)>0.96$,
which are indicated by filled circles. Figure \ref{fig3e}
shows node vs node diagrams for various values of the coupling strength. For $\varepsilon=0.4$
none of the oscillators are synchronized with the drive system (see Fig.~\ref{fig3e}(a)).
Figure \ref{fig3e}(b) indicates that the first seven oscillators are synchronized 
with the drive for $\varepsilon=0.7$. Further increase in the coupling strength results in
increase in the size of the synchronized cluster resulting in the formation of sequential synchronization.
Figures \ref{fig3e}(c) and \ref{fig3e}(d) are depicted for $\varepsilon=0.87$ and $1.1$, respectively,
illustrating sequential synchronization.

To discuss the nature of the synchronization transition locally, we have calculated 
the probability of synchronization for some selected 
systems ($i=15,20$ and $30$) in the array as a function of $\varepsilon$ 
(see Fig.~\ref{fig3}). $\Phi_{i}(\varepsilon)$ of the system
$i = 15$ is plotted as a function of $\varepsilon$ (represented by the filled squares).
In the range of $\varepsilon \in (0, 0.76)$, there
is an absence of any entrainment between the systems resulting in an asynchronous behavior and 
$\Phi_{15}(\varepsilon)$ is practically zero in this region. However,
starting from the value $\varepsilon = 0.76$ and above, there appear some finite values less
than unity attributing to the transition regime. Beyond $\varepsilon = 0.78$, $\Phi_{15}(\varepsilon)$
attains unit value indicating CS. We have also plotted the probability of 
synchronization in Fig.~\ref{fig3} for two more selected systems $i = 20$ and $30$ represented
by the asterisk symbol and filled triangles, respectively, indicating 
the immediate transition to CS locally. The system $i = 20$
attains CS at $\varepsilon = 0.84$ and the system $i = 30$ reaches the CS
state at $\varepsilon = 0.98$. From this figure, one can understand the occurrence of sequential synchronization 
of the individual systems (locally) in the array as a function of the coupling strength. 
To explain global (macroscopic) synchronization phenomenon, we have calculated the average probability of 
synchronization ($\Psi(\varepsilon)$) of the $N=30$ systems as a function of $\varepsilon$ 
and depicted it in Fig.~\ref{fig3} (represented by the filled circles). It
confirms sequential synchronization by gradual increase in $\Psi(\varepsilon)$ as a 
function of $\varepsilon$ (which indeed exactly matches with Fig.~\ref{fig2}).
\begin{figure}
\centering
\includegraphics[width=0.9\columnwidth]{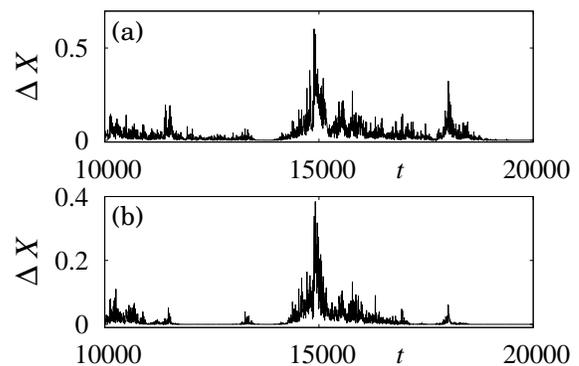}
\caption{\label{fig3c} The average difference ($\Delta X$) of all ($N-1$) 
piecewise linear time-delay systems in the array (Eq.~(\ref{eqn1})) with the drive $x_{1}$ shows an intermittent
synchronization transition. (a) $\varepsilon=0.85$, and (b) $\varepsilon=0.89$.}
\end{figure}

Next, in the transition regime, we observe intermittent synchronization 
in every individual system and this can be characterized qualitatively by a difference 
in the magnitudes of the states between the systems 
($\Delta x_{1,i}=|x_{1}-x_{i}|$,)
for selected ones ($i=15, 20$ and $25$). Figure~\ref{fig3a}(a-c) shows
intermittent synchronization in the above mentioned systems for $\varepsilon=0.76, 0.81$ and $0.86$, respectively. We also find that the synchronization quality in the transition region
depends on the respective positions of the response systems from the drive, as well as on
the distance between the two units in the system and the coupling strength. We have additionally
plotted the difference between the systems $\Delta_{5,20}, \Delta_{5,25}$ and $\Delta_{5,30}$ for the above set of values of coupling strength (Fig.~\ref{fig3a}(d-f)) to demonstrate the above features.  


The statistical features associated with the intermittent dynamics is also analyzed by
the distribution of the laminar phases $\Lambda(t)$ with amplitudes less than a threshold value of
$\Delta$ (here we have choosen $\Delta=\bigl|x_{1}(t)-x_{i}(t)\bigr|=\bigl|0.001\bigr|, \quad i=15,20$ and $25$).
A universal asymptotic power law distribution $\Lambda(t)\propto t^{\alpha}$ is observed for the above threshold
value of $\Delta$ with the exponent $\alpha = -1.5$.
Fig.~\ref{fig3b} shows the laminar phase distribution of the above selected systems. 
The filled circles represent a laminar distribution of the system
$i=15$ for $\varepsilon=0.76$, the filled triangles correspond to the laminar distribution 
of the system $i=20$ for $\varepsilon=0.81$ and the filled squares represent a
laminar distribution of the system $i=30$ for $\varepsilon=0.93$ which clearly display the $-\frac{3}{2}$
power law scaling, a typical characterization of on-off intermittency. It should be noted that this 
result does not change for a large range of $\Delta$.

To understand the phenomenon of intermittent synchronization transition globally in the whole
array, we have calculated the average difference ($\Delta X = \frac{1}{N-1}\sum_{j=2}^{N}|x_{1}-x_{j}|$)
of the ($N-1$) systems in the array with the drive $x_{1}$. Figures \ref{fig3c}(a) 
and \ref{fig3c}(b) show the average difference
for the coupling strengths $\varepsilon=0.85$ and $0.89$, respectively.
\begin{figure}
\centering
\includegraphics[width=0.9\columnwidth]{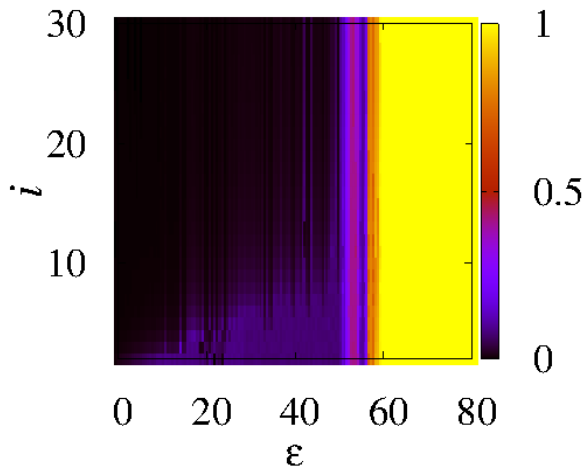}
\caption{\label{fig4a} (Color online) The probability of synchronization $\Phi_{i}(\varepsilon)$ as a function 
of $\varepsilon$ and the system index $i$ illustrating
the occurrence of instantaneous synchronization transition to CS both locally and globally in mutually 
coupled piecewise linear time-delay systems (Eq.~(\ref{eqn4})).}
\end{figure}

The reason behind sequential synchronization transition is in accordance with the 
sequential stabilization of all the unstable periodic orbits of the response systems in the array
as a function of the coupling strength. It is a well established fact that a chaotic/hyperchaotic 
attractor contains an infinite number of UPOs of all periods. Synchronization between
the coupled systems is said to be stable, if all the UPOs of the response systems are stabilized
in the transverse direction to the synchronization manifold. Consequently, all the trajectories
transverse to the synchronization manifold converge to it for suitable values of $\varepsilon$. 
For sequential synchronization, the UPOs in the complex synchronization manifold
of the response systems near to the 
drive are stabilized first for appropriate threshold values of 
the coupling strength $\varepsilon$ as it is increased, while the UPOs of the 
far away systems remain unstable for these values of $\varepsilon$.
Once the coupling is increased
further, the UPOs of the far away systems are gradually stabilized as a function of the
coupling strength. Unfortunately, methods for 
locating UPOs have not been well established for time-delay systems, which has hampered a qualitative 
proof for the gradual stabilization of UPOs by locating them. 
\begin{figure}
\centering
\includegraphics[width=1.0\columnwidth]{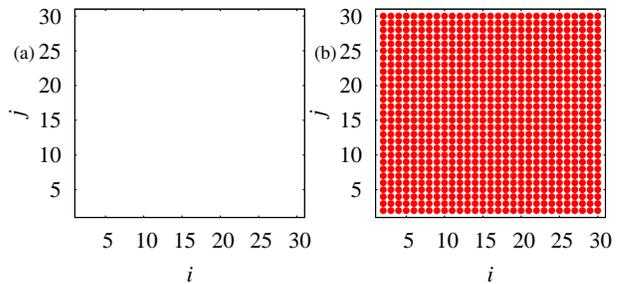}
\caption{\label{fig4b} (Color online) Snap shots of node vs node plots
of mutually coupled piecewise linear systems
indicating instantaneous synchronization.
(a) $\varepsilon=58.0$, (b) $\varepsilon=61.0$.}
\end{figure}
\section{\label{sec:level4}Synchronization in a piecewise linear time-delay systems: Linear array with bidirectional coupling}
In this section, we consider an array of mutually coupled (bidirectional coupling) piecewise linear time-delay systems 
with identical subunits. The dynamical equation then becomes
%
\begin{equation}
\dot{x}_i = -\alpha x_{i}(t)+\beta f(x_{i}(t-\tau))+\varepsilon[x_{i+1}(t)-2x_{i}(t)+x_{i-1}(t)],
\label{eqn4}
\end{equation}
%
where $i = 1, 2, \cdots, N$. We choose open end boundary conditions. 
The parameter values are the same as in Sec.~\ref{sec:level2}. 
The nonlinear function $f(x)$ is chosen as in Eqs.~(\ref{eqn2})-(\ref{eqn3}). 
In the mutual coupling case 
there are no drive and/or response systems where each and every oscillator 
shares the signals mutually with its two nearest neighbors. So the 
synchronization transition is instantaneous due to the mutual sharing
of the signals and one needs a very large value of $\varepsilon$ to attain CS.
In the transition regime, we have observed an intermittent synchronization 
transition in all the systems simultaneously in the array.
\begin{figure}
\centering
\includegraphics[width=0.9\columnwidth]{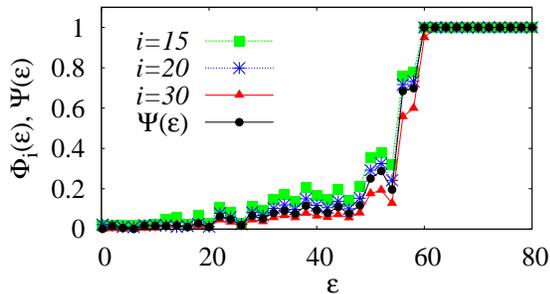}
\caption{\label{fig4} (Color online) The probability of synchronization $\Phi_{i}(\varepsilon)$ of  
selected systems ($i=15,20$ and $30$) and the average probability 
of synchronization ($\Psi(\varepsilon)$) in the mutually coupled array (Eq.~(\ref{eqn4})) as a function 
of $\varepsilon$.}
\end{figure}

We have calculated the probability of synchronization of all the $N=30$ systems in the array
as a function of the coupling strength $\varepsilon$ and the system index $i$ (see Fig.~\ref{fig4a}). 
In this figure the black color represents the desynchronized state ($\Phi(\varepsilon)=0.0$) and 
CS is represented by the yellow color (light gray) ($\Phi(\varepsilon)=1.0$). The transition regime is
indicated by intermediate colors. From this figure one can clearly see that locally every 
individual system requires large values of $\varepsilon$ to attain CS
and globally all the systems synchronize immediately for the same value of $\varepsilon$.

In the mutually coupled array, all the systems get synchronized immediately in a narrow range of the 
coupling strength, in contrast to sequential synchronization. Figure \ref{fig4b}(a) 
is plotted for $\varepsilon=58.0$, where none of the oscillators in the array are  synchronized, 
whereas for $\varepsilon=61.0$, all the systems are completely synchronized as depicted in Fig.~\ref{fig4b}(b).
\begin{figure*}
\centering
\includegraphics[width=2.1\columnwidth]{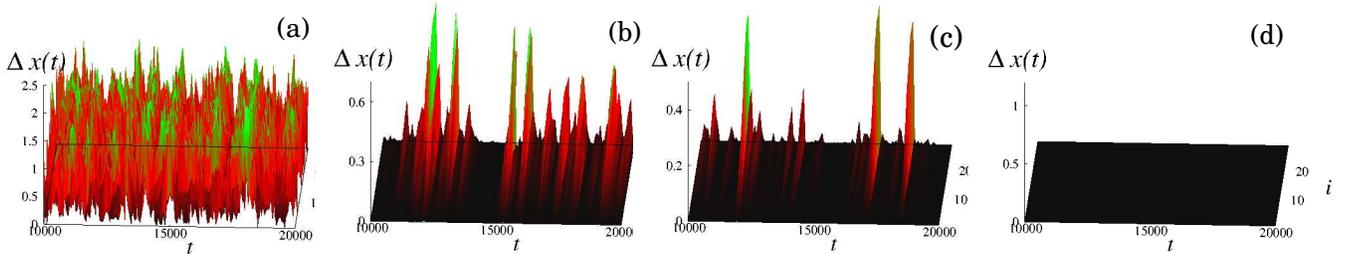}
\caption{\label{fig5} (Color online) The spatiotemporal difference ($\Delta x(t)$) of the 
mutually coupled piecewise linear systems for various values of coupling strengths. 
(a) $\varepsilon = 0.0$, (b) $\varepsilon = 50.2$, (c) $\varepsilon = 54.6$, (d) $\varepsilon = 61.0$. Here the black color
indicates that the difference is zero and the red and green colors (dark and light gray) indicate the bursting amplitudes.}
\end{figure*}

To characterize the nature of synchronization transitions to CS both locally and globally, 
we again use the probability of synchronization $\Phi(\varepsilon)$ and the average probability of synchronization
$\Psi_{i}(\varepsilon)$, respectively. In Fig.~\ref{fig4}, we have plotted
$\Phi_{i}(\varepsilon)$ for some selected piecewise linear
systems ($i = 15, 20, 30$) as a function of $\varepsilon$. 
For instance,  we have illustrated $\Phi_{i}(\varepsilon)$ for the system 
$i = 15$ in Fig.~\ref{fig4}  (represented by the filled squares). 
From this figure, one can observe that in the range of $\varepsilon \in (0, 50)$ there is 
an absence of any entertainment between the systems resulting in asynchronous behavior and 
$\Phi_{15}(\varepsilon)$ is low ($\Phi_{15}(\varepsilon)<0.4$).
However, for $\varepsilon > 50$ there appear oscillations in 
$\Phi_{15}(\varepsilon)$ in the range of 
$\varepsilon \in (50, 60)$ exhibiting intermittent transition. 
Beyond $\varepsilon = 60.0$, $\Phi_{15}(\varepsilon)=1$ 
indicating perfect CS of the system $i = 15$. We have also calculated $\Phi(\varepsilon)$ for the systems 
$i = 20$ and $i = 30$, 
represented by asterisk symbols and filled triangles, respectively, 
which show similar transitions to CS almost at the same value of $\varepsilon$. 
We have also confirmed a similar immediate transition to CS in all the systems
in the array (see Fig.~\ref{fig4a}).  
\begin{figure*}
\centering
\includegraphics[width=2.1\columnwidth]{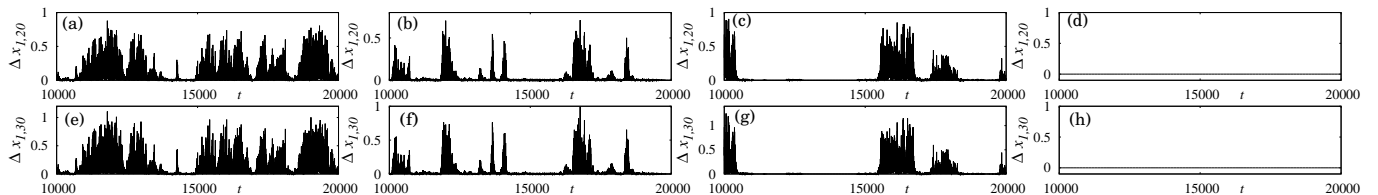}
\caption{\label{fig7} (a)-(d) The difference between systems $1$ and $20$
($\Delta x_{1,20}=\bigl|x_{1}(t)-x_{20}(t)\bigr|$) for $\varepsilon = 50.2$, $54.6$, $58.1$ and $61.0$.
(e)-(h) The difference between systems $1$ and $30$
($\Delta x_{1,30}=\bigl|x_{1}(t)-x_{30}(t)\bigr|$) is plotted for the same set of coupling strength values given above.}
\end{figure*}

To examplify the global synchronization phenomenon, we have calculated the average probability
of synchronization ($\Psi(\epsilon)$) of $N = 30$ systems as a function of the coupling strength
as shown in Fig.~\ref{fig4} by the filled circles.
In the range of $\varepsilon \in (0,54$), there
is an absence of any synchronization and so $\Psi(\epsilon)$ is having zero or low values
($\Psi(\epsilon) < 0.3$). In the range of $\varepsilon \in (54,59$),
$\Psi(\epsilon)$ is characterized by some finite values less than unity and beyond $\varepsilon > 59.0$
there is a sudden jump to the value of $\Psi(\epsilon)=1.0$ corrobrating all the systems 
synchronize immediately at the same value of the coupling strength attributing to the occurrence of global CS. 
Further, in the transition region we have found long time intervals of high quality synchronization
which is interrupted at irregular time intervals by intermittent chaotic bursts simultaneously in all the systems 
in the array which we call as GIS.

To demonstrate the existence of GIS, we have 
calculated the spatiotemporal
difference ($\Delta x(t) = \bigl|x_{1}(t)-x_{i}(t)\bigr|, \quad i = 2, 3, \cdots, 30$) of the array 
as a function of time and the oscillator index $i$ as in
Fig.~\ref{fig5} for different values of $\varepsilon$. Here the black color indicates 
zero difference ($\Delta x(t)=0.0$) and the red and green colors (dark and light gray) indicate bursting amplitudes. 
In the absence of coupling
($\varepsilon = 0.0$), the systems are evolving independently and so
there is no correlation between the systems as shown in Fig.~\ref{fig5}(a). If we increase the coupling, 
we observe several intermittent bursts along with the 
synchronization as depicted in Figs.~\ref{fig5}(b) and \ref{fig5}(c) for
$\varepsilon = 50.2,$ and $54.6$, respectively. From these figures, one can clearly see 
the occurrence of aperiodic intermittent chaotic bursts along with the synchronized regions simultaneously 
in all the systems in the array. Beyond $\varepsilon > 60$ one can observe CS as
illustrated in Fig.~\ref{fig5}(d) where the spatiotemporal difference of
the systems is exactly zero for $\varepsilon = 61.0$. 

To elaborate the occurrence of GIS in the 
array more clearly, we have calculated the difference between the systems coupled in the array and plotted  
for some selected systems ($i = 20$ and $30$) for different values of the coupling strength.
The difference between the systems $1$ and $20$, $\Delta x_{1,20}(t) = \bigl|x_{1}(t)-x_{20}(t)\bigr|$, is plotted for
$\varepsilon = 50.2, 54.6$ and $58.1$ in 
Figs.~\ref{fig7}(a) - \ref{fig7}(c), respectively, which clearly
displays the existence of aperiodic intermittent bursts along with the synchronized regions. We have also 
plotted the difference $\Delta x_{1,30}(t) = \bigl|x_{1}(t)-x_{30}(t)\bigr|$ for the same values of 
$\varepsilon$ as shown in Figs.~\ref{fig7}(e) - \ref{fig7}(g). It is to be noted that in both
systems ($i = 20, 30$) the intermittent bursts simultaneously occur at the same time and this occurs 
in all the other systems connected in the array
confirming the existence of GIS. In Figs.~\ref{fig7}(d)
and \ref{fig7}(h) the difference between the systems completely vanish for $\varepsilon = 61.0$
indicating the occurrence of CS. We have plotted the above figures with $10^{4}$ time units after
leaving a sufficient number of transients. 

\begin{figure}
\centering
\includegraphics[width=1.0\columnwidth]{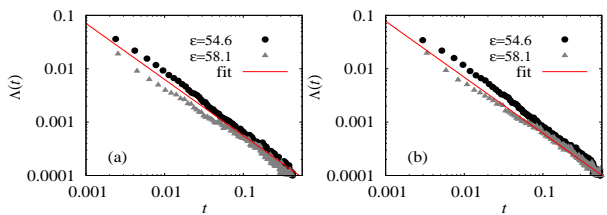}
\caption{\label{fig8} (color online) The statistical distribution of the laminar phase for
selected piecewise linear systems (Eq.~\ref{eqn4}) (a)$i = 20$ and (b) $i = 30$ satisfying a $-\frac{3}{2}$ power law scaling for 
the coupling strength $\varepsilon = 54.6 $ and $58.1$.}
\end{figure}

Further statistical features associated with the intermittent dynamics of the entire array are also analyzed by calculating
the distribution of the laminar phases $\Lambda(t)$, which is shown in Figs.~\ref{fig8}(a) 
and \ref{fig8}(b) for selected systems $i=20$ and $30$, respectively, for $\varepsilon = 54.6$ 
and $58.1$ which clearly display the $-\frac{3}{2}$ power law scaling to confirm the on-off
intermittency.

The reason for the occurrence of GIS can be explained
as follows: As we have already explained, a chaotic attractor can be considered as a pool of
infinitely many UPOs of all periods. Synchronization between the systems are asymptotically stable,
if all the UPOs of the systems are stabilized in the transverse direction to the synchronization 
manifold. Consequently, all the trajectories transverse to the synchronization manifold converge 
to it for suitable values of the coupling strength and this is reflected in the stabilization of the UPOs
upon synchronization. From our results, we find that the UPOs of the systems are 
stabilized in the complex synchronization manifold only for a very large value of coupling strength 
after a certain threshold value.
It is also to be noted that the intermittency 
transition in the case of a bidirectional coupling configuration is due to the fact that the strength 
of the coupling $\varepsilon$ contributes only less significantly to stabilize the UPOs as the error in the coupling
term in Eq.~(\ref{eqn4}) gradually becomes smaller from the transition regime after a certain threshold value
of the coupling strength.

\section{\label{sec:level8}Conclusion}
In conclusion, we have shown the existence of sequential and instantaneous synchronization transitions
in an array of time-delay systems with different coupling configurations. If the
systems are coupled with unidirectional configuration, we have observed an immediate synchronization transition to
CS microscopically and if we consider the 
macroscopic synchronization behavior of the entire array we find that
the transition region is gradually increasing as a function of $\varepsilon$
due to sequential synchronization which is verified by the probability of synchronization and
average probability of synchronization. In the transition regime we have observed the 
existence of intermittent synchronization. On the other hand, if we
consider an array of mutually coupled time-delay systems, every individual system 
(microscopically) synchronizes immediately
for a very large value of $\varepsilon$ and globally (macroscopically) the synchronization 
transition occurs immediately in the whole array. In the transition region a new type of synchronization
called GIS occurs which is characterized by long intervals of high quality synchronization interrupted at
irregular times by intermittent chaotic bursts simultaneously in all the systems.

The reason (mechanism) for these two distinct transition scenarios is explained based on unstable periodic
orbit theory. The GIS is confirmed using the 
spatiotemporal difference and a power law behavior 
of the laminar length distributions with 
$-\frac{3}{2}$ power law scaling.  The above studies have been carried out in 
a well known piecewise linear time-delay system. We have also confirmed the occurrence of the above
results for another well known time-delay systems namely the Mackey-Glass system \cite{mcmlg1977} with an 
array length of $N=50$ and we do observe
the same kind of sequential and instantaneous synchronization transitions preceded by GIS for unidirectional and
bidirectional coupling configurations, respectively.

\begin{acknowledgments}
The work of R.~S. and M.~L. is supported by a Department of Science and
Technology (DST), IRHPA research project.
M.~L. is also supported by a Department of Atomic Energy Raja Ramanna fellowship and a DST Ramanna program.
D.~V.~S. and J. K. acknowledge the support from  EU  under project No. 240763 PHOCUS(FP7-ICT-2009-C) and 
J. K. acknowledges the support from IRTG 1740(DFG).
\end{acknowledgments}


\end{document}